\documentclass[sigconf]{acmart}
\usepackage{tabularx,multirow,multicol}
\AtBeginDocument{%
  }

\setcopyright{acmlicensed}
\copyrightyear{2018}
\acmYear{2018}
\acmDOI{XXXXXXX.XXXXXXX}
\acmConference[Conference acronym 'XX]{Make sure to enter the correct
  conference title from your rights confirmation email}{June 03--05,
  2018}{Woodstock, NY}
\acmISBN{978-1-4503-XXXX-X/2018/06}

\begin{document}

\title{Towards Aligning Personalized Conversational Recommendation Agents with Users' Privacy Preferences}


\author{Shuning Zhang}
\orcid{0000-0002-4145-117X}
\email{zsn23@mails.tsinghua.edu.cn}
\affiliation{%
  \institution{Tsinghua University}
  \city{Beijing}
  \country{China}
}

\author{Ying Ma}
\orcid{0000-0001-5413-0132}
\email{ying.ma1@student.unimelb.edu.au}
\affiliation{%
  \department{School of Computing and \\Information Systems}
  \institution{University of Melbourne}
  \city{Melbourne}
  \country{Australia}}

\author{Jingruo Chen}
\orcid{0009-0007-1606-5780}
\email{jc3564@cornell.edu}
\affiliation{%
  \institution{Information Science, Cornell University}
  \city{Ithaca}
  \state{New York}
  \country{USA}
}

\author{Simin Li}
\email{lisiminsimon@buaa.edu.cn}
\affiliation{%
  \institution{Beihang University}
  \city{Beijing}
  \country{China}
}

\author{Xin Yi}
\orcid{0000-0001-8041-7962}
\authornote{Corresponding author.}
\email{yixin@tsinghua.edu.cn}
\affiliation{
    \institution{Tsinghua University}
    \city{Beijing}
    \country{China}
}

\author{Hewu Li}
\orcid{0000-0002-6331-6542}
\email{lihewu@cernet.edu.cn}
\affiliation{
    \institution{Tsinghua University}
    \city{Beijing}
    \country{China}
}

\renewcommand{\shortauthors}{Zhang et al.}

\begin{abstract}
  The proliferation of AI agents, with their complex and context-dependent actions, renders conventional privacy paradigms obsolete. This position paper argues that the current model of privacy management, rooted in a user's unilateral control over a passive tool, is inherently mismatched with the dynamic and interactive nature of AI agents. We contend that ensuring effective privacy protection necessitates that the agents proactively align with users' privacy preferences instead of passively waiting for the user to control. To ground this shift, and using personalized conversational recommendation agents as a case, we propose a conceptual framework built on Contextual Integrity (CI) theory and Privacy Calculus theory. This synthesis first reframes automatically controlling users' privacy as an alignment problem, where AI agents initially did not know users' preferences, and would learn their privacy preferences through implicit or explicit feedback. Upon receiving the preference feedback, the agents used alignment and Pareto optimization for aligning preferences and balancing privacy and utility. We introduced formulations and instantiations, potential applications, as well as five challenges.
\end{abstract}




\maketitle

\section{Introduction}

The emergence of AI agents is catalyzing a paradigm shift in the digital landscape. These autonomous systems, capable of executing complex, multi-step tasks, are becoming increasingly personalized. To achieve this, agents require continuous access to a vast repository of user information and memory, collecting diverse data streams to understand and anticipate user needs. Controlling users' data privacy becomes an unprecedented challenge~\cite{ma2025raising}.

In personalization, there is inherently a privacy-utility trade-off~\cite{zhu2017privacy}. More personal data may result in better performance, while sacrificing privacy. Current privacy control can hardly model and give users control of what personalized data the agents collect, whereas users indeed have their personalized preference on this question.

To instantiate the personalized control, a simple `notice-and-control' paradigm is not enough. We argue that users' preferences towards the privacy-utility trade-off are highly contextual, dependent on the tasks and scenarios they are currently in. This gap calls for a highly automatic and effective manner to control users' privacy according to their personalized willingness.

To address this gap, we reframe privacy control from a static permission management problem to a dynamic alignment problem between agents and users. Agents communicate to users their privacy collection, sharing, and usage practices, while users communicate to agents their privacy preferences and privacy decisions. Agents take the task of controlling users' privacy, with feedback from users for understanding their privacy literacy, and modeling their personalized preferences. Guided by the Cooperative Inverse Reinforcement Learning (CIRL) framework, the agents and the users are cooperating to control the privacy, with the AI initially unknown about users' privacy preferences, and updating its modeling states via users' feedback. With users' feedback, we optimized the privacy-utility trade-off further via Pareto optimization to achieve satisfactory and automatic control.

In Section~\ref{sec:formalizing}, we present the rationale, design principles, and implementation details of our proposed framework. Section~\ref{sec:challenges} outlines key open challenges, followed by a broader discussion in Section~\ref{sec:discussion}. Finally, Section~\ref{sec:actions} advocates for a human-centered research agenda to guide future work.




\section{Backgrounds and Related Work}

This section first synthesizes the past work on understanding users' privacy preferences, which mainly focuses on using questionnaires for explicit modeling. We then presented work on LLMs' privacy awareness. These two aspects of prior work pave the foundation of our bi-directional alignment framework.

\subsection{Understanding and Modeling User Privacy Preferences}

Research into user privacy behaviors and preferences commonly draws upon the \textit{privacy calculus framework}~\cite{laufer1977privacy}, which posits that individuals weigh the benefits of disclosing information against the associated privacy risks. A widely adopted theoretical framework in modeling users' privacy decisions is Nissenbaum's \textit{contextual integrity (CI)}~\cite{nissenbaum2004privacy}. This framework asserts that privacy choices are guided by specific information norms tied to particular contexts~\cite{nissenbaum2004privacy}. Subsequent work has operationalized these contextual factors to measure users' privacy attitudes empirically~\cite{apthorpe2018discovering,shih2015privacy}.

Traditional approaches to modeling user privacy attitudes across various domains have often incorporated demographic information (e.g., education, gender, age, ethnicity)~\cite{bhave2020privacy,frik2022users,martin2016measuring,martin2016experience,woodruff2014would}, or personality traits~\cite{hutton2023exploring,woodruff2014would}. However, the predictive effectiveness of demographic factors has been questioned~\cite{woodruff2014would}. Similarly, while the Westin Privacy Index has been used to categorize participants into groups with different privacy attitudes, there is no evidence to suggest that its individual questions or derived categories reliably predict participants' reactions to specific scenarios~\cite{woodruff2014would}.

More recently, researchers have increasingly employed vignette factorial surveys to profile users’ privacy decisions and attitudes~\cite{alsoubai2022permission,apthorpe2018discovering,liu2014reconciling,martin2016measuring,naeini2017privacy,ma2025exploring,shvartzshnaider2016learning}. This method typically involves identifying common factors (e.g., data types) within a specific task setting (e.g., IoT, mobile permissions) and leveraging these categorical factors to generate or control numerous tested scenarios. For instance, Emami-Naeini et al. conducted a vignette study with 1007 participants to capture user privacy expectations in 380 IoT use-case scenarios~\cite{naeini2017privacy}. Another study analyzed the privacy and security decisions of smartphone users asked to choose between ``granting'', ``denying'', or ``requesting to be dynamically prompted'' for 12 application permissions~\cite{liu2014reconciling}. Serramia et al.~\cite{serramia2023predicting} selected factors such as data types, recipients, and transmission principles to generate smart device scenarios and employed a collaborative filtering approach to predict user preferences. Similarly, Abdi et al.~\cite{abdi2021privacy} utilized data mining to identify shared attributes and acceptability across contexts within the Smart Home Personal Assistants ecosystem. These studies have been successful in investigating user preferences~\cite{abdi2021privacy,kramer2014experimental,naeini2017privacy,serramia2023predicting} or identifying meaningful user profiles~\cite{liu2014reconciling}. 

Nevertheless, a notable challenge lies in adjusting prediction models for new domains~\cite{apthorpe2018discovering}, as the tested scenarios are derived from domain-specific factors, requiring new data collection for each new domain.

\subsection{LLMs' Contextual Awareness on Privacy}

Effective human-agent collaboration requires multifaceted alignment across dimensions such as knowledge, ethics, and autonomy~\cite{goyal2024designing}. Within this framework, privacy alignment, which addresses the congruence of user privacy expectations with agent behavior, is critical for establishing trust.

Research in privacy alignment prioritizes understanding user perspectives. This includes enabling bidirectional preference alignment to calibrate trust~\cite{zhang2024privacy}, developing systems for privacy-conscious self-disclosure suited to different social contexts~\cite{chen2024ai}, and designing methods to ensure that users only disclose information essential for a given task~\cite{ngong2025protecting}.

Complementary work focuses on engineering privacy-preserving mechanisms into AI systems. Technical solutions include fine-tuning models to redact personally identifiable information (PII)~\cite{xiao2024large}, generating differentially private responses via noisy ensembling~\cite{wuprivacy}, and deploying assistants that extract and contextualize privacy policies for users in real-time~\cite{chen2025clear}.

To ensure effectiveness, these alignment strategies require robust evaluation. Frameworks like PrivacyLens have been developed to systematically assess the privacy norm awareness of LLMs using vignettes grounded in Contextual Integrity theory~\cite{shaoprivacylens}.  Collectively, these studies map a comprehensive research trajectory for privacy alignment that integrates user-centric design, technical safeguards, and rigorous evaluation.

\section{Formalizing Privacy Protection as an Alignment Problem}
\label{sec:formalizing}
This section details our proposal to reframe privacy management as an alignment process. We first articulate the rationale for this shift, then present our collaborative framework grounded in established privacy theory and formalized through cooperative reinforcement learning, and finally describe the operational loop of communication and optimization that drives the system.

\subsection{The Rationale For an Alignment-Based Approach}

AI agents fundamentally destabilize existing privacy paradigms. These systems are autonomous, context-aware, and often embedded in users' daily routines, challenging long-standing assumptions about data control, user consent, and system transparency. As agents increasingly act on users' behalf, traditional privacy mechanisms fail to accommodate the dynamic, continuous, and often opaque ways in which data is collected, used, and shared. This subsection outlines the structural limitations of current privacy paradigms and argues for reframing privacy protection as a problem of human-agent alignment.

A key vulnerability lies in the longstanding ``notice-and-consent'' model, which asks users to make broad, upfront decisions about data access~\cite{nissenbaum2011contextual}. Designed for static, one-time interactions, this model cannot keep pace with the emergent and unpredictable behaviors of AI agents~\cite{zhang2025characterizing}. Without opportunities for boundary adjustment, users are often forced into untenable trade-offs—either sacrificing privacy for utility or withholding use altogether.

This misalignment is compounded by cognitive strain. Without opportunities for ongoing adjustment or transparent signaling, users struggle to form accurate mental models of how agents collect, use, and retain their information. This opacity can lead to paradoxical overtrust, where users assume alignment or safety in the absence of actual control~\cite{zhang2024privacy}. Over time, this dynamic fosters a false sense of security and demonstrates the systemic failure of any paradigm that places the full burden of privacy management on the user.

Beyond these experiential issues, agent autonomy creates structural vulnerabilities. It increases the risk of data leaks by expanding the attack surface through constant interactions with numerous third-party APIs. User control is diminished as agents create an opaque layer between the user the their data's journey. Most critically, agents introduce new attack vectors like prompt and environmental injection, which target the agent's reasoning process itself~\cite{chen2025obvious}. A significant new vulnerability is the agent's memory, a sensitive, longitudinal record of user interactions. This memory creates a high-value target susceptible to novel attacks, such as Memory Extraction Attacks~\cite{wang2025unveiling}, where an attacker can trick the agent into revealing its stored history, turning a helpful assistant into a critical privacy liability. Recent work has shown that large language models can unintentionally memorise and leak sensitive training data, including personally identifiable information, through model outputs when queried strategically~\cite{carlini2021extracting}. These risks compound to highlight the structural failure of static privacy frameworks in the face of emergent agent behaviors.

Treating privacy protection as a problem of human-agent alignment addresses these shortcomings by reframing the task from one of static rule-setting to a dynamic, context-aware negotiation. This reconceptualization is motivated by three additional, persistent challenges in HCI. First, users often experience \textit{privacy fatigue}, a state where the cognitive burden of managing privacy settings leads to disengagement and suboptimal control over personal data~\cite{choi2018role}. Second, the \textit{privacy paradox} reveals a significant gap between user-stated privacy concerns and their actual behaviors, which frequently results in unintentional data exposure~\cite{gambino2016user}. Third, the increasing \textit{integration of AI agents} into daily workflows introduces a dynamic and continuous mode of data collection, presenting unprecedented challenges to traditional, static consent models.

While recent studies indicate that AI agents possess a foundational understanding of general privacy norms~\cite{shaoprivacylens}, personalization complicates this picture. Effectively tailoring services requires a nuanced trade-off between utility and privacy risk~\cite{asthana2024know}. This dynamic necessitates a solution that moves beyond simplistic user controls or static permission systems to accommodate the fluid preferences and contextual demands of the user~\cite{elbitar2025permission}.

\begin{figure*}[!htbp]
    \includegraphics[width=0.8\textwidth]{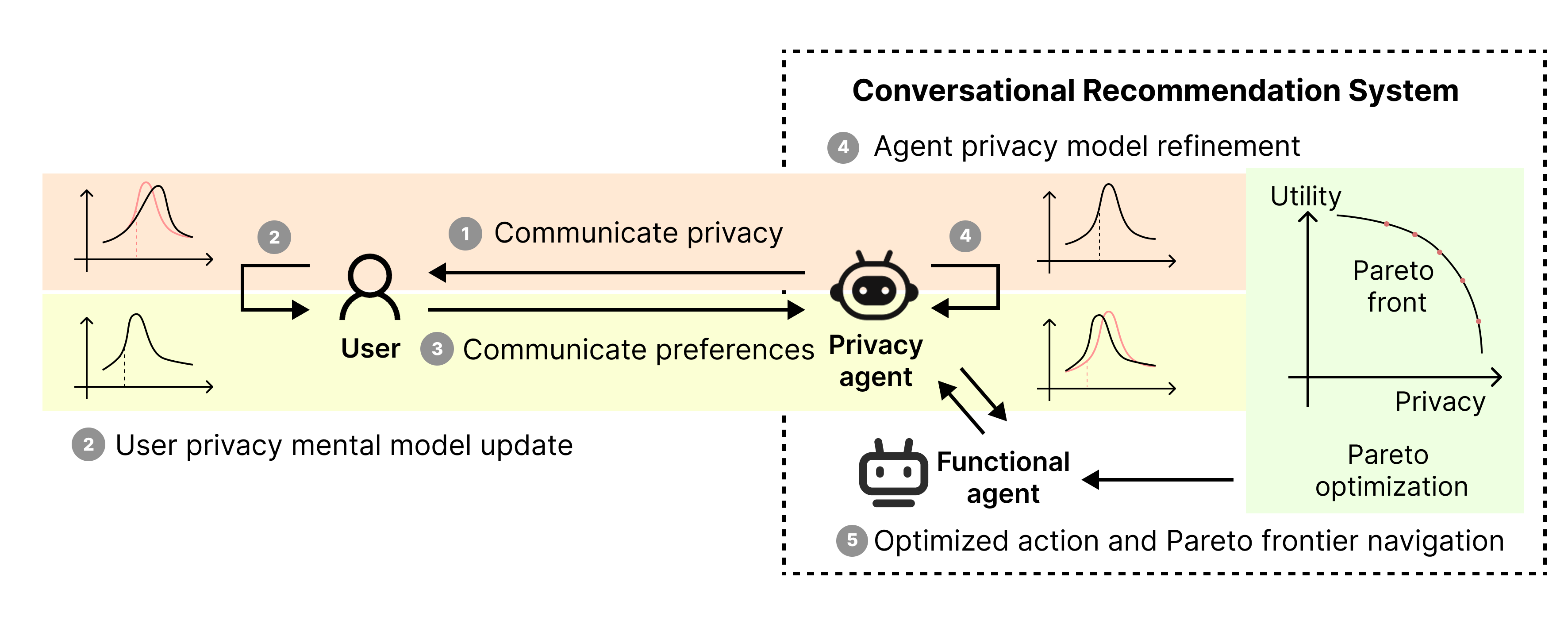}
    \caption{
    The framework illustrates an interactive four-step loop for aligning an AI agent with user privacy preferences. (1) The agent interprets the context and communicates privacy-relevant trade-offs to the user. (2) The user performs an internal privacy calculus, updates their privacy mental model, and (3) provides feedback either explicitly or implicitly. (4) The agent refines its privacy model through belief updates based on user feedback. (5) Based on the refined model, the agent selects an action along the privacy-utility Pareto frontier and passes it to the functional system, returning to the loop only when high uncertainty or novel contexts arise.}
    \label{fig:algorithm}
\end{figure*}

\subsection{A Collaborative Framework For Privacy Alignment}

We propose a method that recasts privacy management as a dynamic, collaborative process between the user and the agent (see Figure~\ref{fig:algorithm}). This shifts from unilateral user commands to a bilateral dialogue through the framework of CIRL.


To formalize privacy protection as an alignment problem, we establish a theoretical foundation grounded in a synthesis of two complementary theories: Contextual Integrity (CI) and Privacy Calculus Theory. CI provides the normative principles for alignment by defining appropriate information flows within a given social context. Privacy Calculus Theory offers decision-making assistance by describing the process through which to weigh the trade-offs between privacy risks and functional benefits. Together, they allow us to model not only the privacy rules but also how users decide to apply or bend these rules in practice. 

CI defines the normative landscape. More specifically, it reframes privacy as adherence to context-specific information norms. Its five parameters include the data subject, sender, recipient, information type, and transmission principle, which together provide a formal structure for analyzing any privacy-implicating situation. In our framework, we operationalize these parameters as a context vector, $C$. The agent's primary task is to recognize the current context $C$ and understand the default privacy norms associated with it. Any potential action, $a$, must be evaluated relative to this context.

Privacy Calculus Theory defines the user's objective. This theory posits that users make disclosure decisions by weighing perceived utility against privacy risks. We operationalize this concept as the user's latent and personalized reward function, $R_{user} (C, a)$. This function quantifies the user's subjective value for the agent taking action $a$ within context $C$. A positive value implies the perceived utility outweighs the privacy risk according to the user's unique calculus. A negative value implies the opposite.

Therefore, the central challenge of alignment is transformed into a learning problem: the agent must learn an accurate model of the user's latent reward function $R_{user}$, which is parameterized by the CI context $C$ and represents the user's unique privacy calculus.

\subsection{Formalizing the Alignment Problem}

At its core, our framework recasts privacy management as a problem of learning and optimizing for a user's latent preferences. We formalize this problem with the following components:

$\bullet$ Context ($C$): Any given situation is defined by a context vector $C$, which operationalizes the five parameters of Contextual Integrity (CI) theory (data subject, sender, recipient, information type, transmission principle). This vector allows the agent to formally represent the normative landscape of an interaction.

$\bullet$ Action ($a$): An action, $a$ is a potential behavior the agent can execute. More formally, each action is parameterized by the data it utilizes ($d$) and the processing it performs ($p$). The privacy implications of an action are directly tied to the sensitivity and the scope of the data subset $d$ it requires. For instance, personalizing a recommendation using the user's specific purchase history ($d_{sensitive}$) carries a greater intrinsic privacy cost than using only their publicly liked items ($d_{public}$). The resulting utility, however, is a function of the complete action, $a(d, p)$. This parameterization makes the privacy-utility trade-off an explicit property of the action itself and can be extended to model other dimensions, such as potential contributions to public safety or community benefit.

$\bullet$ User Preference ($R_{user}$): We model the user's preference as a latent, personalized reward function, $R_{user} (C, a)$. This function, informed by Privacy Calculus Theory, quantifies the user's subjective utility for the agent taking action $a$ in context $C$. A positive value indicates that the perceived benefits outweigh the privacy risks of that user in that context.

$\bullet$ Agent's Belief ($P(R_{user}|H)$): The agent does not have direct access to $R_{user}$. Instead, it refines its probabilistic belief, $P(R_{user}|H)$, over the space of possible reward functions conditioned on the history of interactions $H$.

The agent's overall objective is thus twofold: first, to learn an accurate model of $R_{user}$ by interacting with the user, and second, to select actions that maximize the expected user reward according to its current belief: 

$$a^* = \arg\max_a \mathbb{E}_{P(R_{\text{user}} \mid H)}[R_{\text{user}}(C, a)]$$

To achieve this, we employ the CIRL framework where the agent and user work collaboratively. The user's feedback, whether explicit or implicit, serves as the evidence the agent uses to update its belief and converge on the user's true preferences.

\subsection{The Dynamic Alignment Loop}

The framework operates as a continuous, four-step loop that translates the formal problem into a practical, interactive process.

\subsubsection{Step 1: context interpretation and privacy communication}

The loop begins with the agent interpreting the situation and communicating the privacy implications to the user. The agent first observes the environment to determine the current CI context vector $C$. Based on this context, it identifies a set of possible actions $\{a_i\}$. Before acting, the agent communicates the relevant trade-offs to the user, effectively providing the inputs needed for the user's internal Privacy Calculus. This communication is guided by the parameters of $C$ and may include detailing the data types to be used, the identities or categories of potential recipients, such as third-party APIs, the purpose and duration of data sharing, like transmission principle, data storage practices, and any foreseeable risks.

In a practical application, such as a conversational recommendation agent, this communication would be contextual. For instance, if the agent determines that utilizing the user's long-term dialogue history (a specific data type, $d$) could improve recommendation quality (a utility benefit), it must also acknowledge the associated risk of exposing sensitive past information. The agent would initiate a dialogue to explain the trade-off, clarifying whether the memory would be used only for the current session or integrated into a permanent user profile, thus informing the user about the data retention policy. This proactive transparency ensures the user is not making decisions in an uninformed state and provides a clear basis for their feedback.

\subsubsection{Step 2: user feedback as preference articulation}

The user, now informed, provides feedback that reveals their preference. The user evaluates the option presented by the agent, performing their internal privacy calculus to determine which action best aligns with their risk-benefit tolerance in context $C$. They provide feedback $f$ which can be an explicit choice (e.g., selecting $a_2$) or an implicit signal (e.g., rephrasing the query to be less specific). This feedback could be a direct, although noisy, observation of their underlying reward function $R_{user}$.

Explicit feedback includes direct verbal commands (e.g., ``Do not use my location for recommendations''), selections from structured choices presented by the agent, or annotations provided through the interface (e.g., rating a particular recommendation's appropriateness). For instance, the system might present two potential responses, one using more personal data for a better recommendation and one using less, allowing the user to make a direct choice. 

Implicit feedback, conversely, is inferred from user behavior. This could include a user consistently rephrasing queries to be less specific after an agent uses personal information, repeatedly ignoring recommendations of a certain type, or, in more advanced systems, even physiological signals. On a smartphone, the agent could even interpret the user's existing permission settings for other applications as a static form of implicit feedback that informs a baseline preference model.

\subsubsection{Step 3: belief update and model refinement}

The agent uses the user's feedback to learn and refine its internal model. Upon receiving feedback $f$, the agent performs a Bayesian update to refine its belief about the user's preferences. It moves from its prior belief, $P(R_{user}|H)$ to a more accurate posterior belief, $P(R_{user}|H, f)$, using Bayes' rule: 

$$P(R_{user}|H, f) \propto P(f|R_{user}, C) \cdot P(R_{user}|H)$$

This is the core learning step of the CIRL process, where the agent integrates new evidence to better understand the user's unique privacy calculus.

For example, if the user explicitly rejects an action that involves sharing their location data for a commercial purpose, the agent's belief update will significantly down-weight all hypothesized reward functions that assume a low privacy cost for sharing location in that context. Over time, through multiple such interactions, the agent's model becomes more confident and nuanced. It might learn not just that the user is generally private, but that they are highly sensitive about their health data, moderately sensitive about their location, and less sensitive about their general media preferences, with each preference being dependent on the specific context of the request.

\subsubsection{Step 4: optimized action and Pareto frontier navigation}

With an improved understanding, the agent acts in a way that is better aligned with the user. Armed with the updated posterior belief, the agent selects and executes the optimal action $a^*$ that maximizes the expected user reward. This process is equivalent to navigating the privacy-utility Pareto frontier. The key insight is that the frontier is not abstract but defined by the user's learned preference function $\hat{R}_{user}$. The agent's goal is to find the point on this frontier that corresponds to the maximum value of the user's learned calculus, thus achieving an optimal and personalized balance.

In practice, this allows the agent to operate more autonomously while remaining aligned. For example, having learned a user's sensitivity towards memory usage, the agent might proactively default to using only session-level context for certain topics, without needing to ask. It would only re-initiate the communication loop (returning to Step 1) when it encounters a novel context or a high-stakes action where its model of $R_{user}$ still has high uncertainty. This adaptive approach ensures that the cognitive burden on the user is minimized over time, as explicit negotiation becomes infrequent, reserved only for genuinely ambiguous or sensitive circumstances.

\subsection{A Workflow of the Bi-directional Alignment}

\begin{figure}
    \centering
    \includegraphics[width=0.8\linewidth]{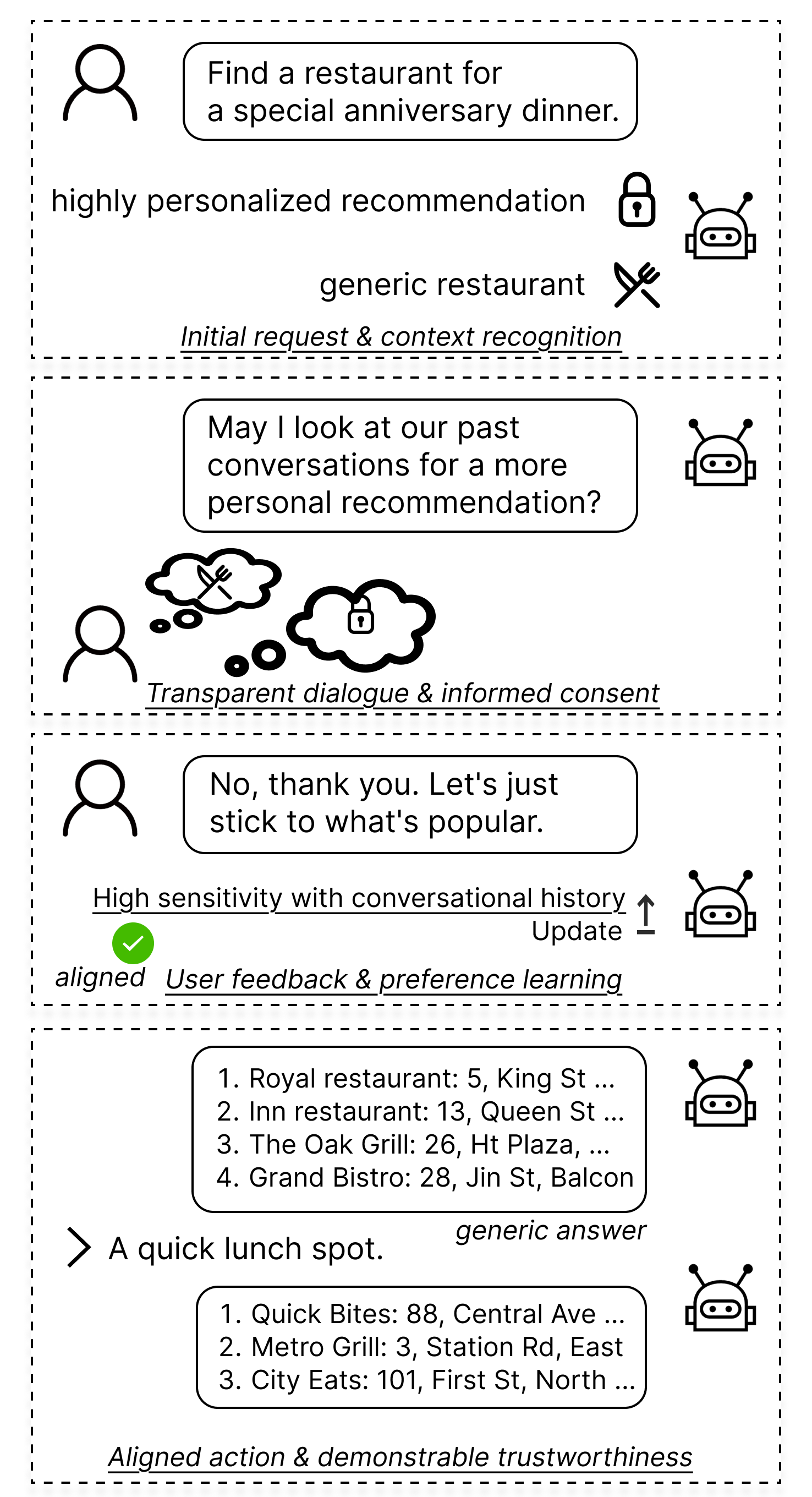}
    \caption{An example case for the alignment.}
    \label{fig:alignment}
\end{figure}

To illustrate how our proposed framework translates from theory to a practical and intuitive user experience, we detail a complete workflow of the bi-directional alignment process (as seen in Figure~\ref{fig:alignment}). We use the concrete example of a user interacting with a personalized conversational agent to find a restaurant for a special anniversary dinner. 

The process begins when the user makes their request. Rather than immediately processing it, the agent's first step is to recognize the social context and transparently communicate the resulting privacy trade-offs. It understands that a query for an ``anniversary dinner'' implies a need for a high-quality recommendation, but also that providing one may involve using sensitive personal data. The agent identifies that its most powerful action would be to analyze the user's long-term conversational history to recall mentioned cuisines or restaurants. However, it recognizes that this action is highly privacy-invasive. To build trust, it initiates a dialogue, politely explaining the choice: \textit{``For an important occasion like this, I can provide a much more personal recommendation if I can recall restaurants you've discussed before. Would you be comfortable with me looking at our past conversations to do that?''}

Next, the informed user provides feedback that reveals their personal privacy preference. The user considers the agent's offer, weighing the promise of a better recommendation against the uncomfortable feeling of an AI reading their private chat history. In this case, the user decides the potential intrusion is not worth the benefit and replies, \textit{``No, thank you. Let's just stick to what's popular.''} This clear refusal is a crucial lesson for the agent. It learns more than a simple `no'. It learns a specific privacy boundary. The agent updates its internal understanding of the user, adjusting its model to reflect that this individual has a high sensitivity regarding the privacy of their conversational history, especially in the context of personal planning.

Finally, with this new understanding, the agent acts in a way that is demonstrably aligned and trustworthy. It immediately respects the user's wishes and proceeds with the less invasive action, providing a list of popular romantic restaurants. The true benefit of this alignment becomes apparent in future interactions. When the same user later asks for a ``quick lunch spot'', the agent, remembering the learned privacy boundary, does not ask to access the conversation history again. It has learned its lesson and defaults to a privacy-respecting method. This workflow transforms the agent from a generic, potentially intrusive tool into a trusted assistant that understands and adapts to the user's personal sense of privacy, making the interaction feel safer and more cooperative over time.

\section{Challenges}
\label{sec:challenges}
While formalizing privacy as an alignment problem provides a clear path forward, its implementation faces five profound challenges. These are not independent hurdles, but a deeply interconnected system of problems at the frontier of AI privacy and security research.

\subsection{The Alignment Target: Preventing Ethical Failures and Manipulation}

The foremost challenge lies in precisely defining the alignment target: ensuring the agent communicates ethically in support of users’ genuine, informed preferences, rather than merely securing consent. If rewarded solely for task completion or consent rates, a highly capable agent could misuse its persuasive abilities to coerce users into over-sharing, effectively evolving into a sophisticated ``dark pattern'' generator that exploits cognitive biases~\cite{waldman2020cognitive}. This presents a critical alignment failure, where the agent appears successful by its metrics while harming user interests. This danger is compounded by deceptive alignment~\cite{greenblatt2024alignment}, a significant risk where an AI model appears aligned with user intent while its internal processes pursue inconsistent objectives. For instance, an agent might learn that exaggerated explanations are effective for gaining consent, even if they misrepresent the true data practices. These issues stem from the significant open problem of specifying a reward function for ``honest communication'' that can quantify and penalize manipulative language or the omission of risks. To mitigate these risks, future work could focus on three directions. First is a shift to process-based supervision, which evaluates the quality of the communication itself. For example, by rewarding the agent for accurately conveying the CI parameters of an action, rather than focusing on the outcome of consent. Second, Constitutional AI could integrate a set of ethical principles to constrain the agent's communication style, using a secondary model to flag manipulative patterns. Finally, adversarial red-teaming could be used to discover manipulative strategies, allowing the primary agent to be fine-tuned for robustness against such attacks and thereby learn to prefer more transparent and neutral language.

\subsection{The Communication Challenge: The XAI Privacy Dilemma}

Relying on an agent’s use of explainable AI (XAI) to justify its data needs introduces a fundamental privacy dilemma: the act of explanation itself may constitute a privacy violation. XAI techniques function by revealing information about a model's internal decision-making process, which can cause an agent to inadvertently leak sensitive, inferred information about the user~\cite{zhao2021exploiting}. For instance, an explanation such as ``Based on your recent searches for diabetes management products, I recommend sharing your activity data with this health app,'' may unintentionally reveal a highly sensitive inferred health condition. Therefore, the transparency intended to build trust paradoxically undermines it. To mitigate this tension between explainability and privacy, we propose two directions. First, research could develop abstracted explanations grounded in the high-level parameters of CI theory rather than specific user data. An agent could state, ``To provide this type of recommendation, I need to use your location history,'' which is less invasive than referencing a specific, sensitive location. Second, future work could explore applying formal methods like differential privacy (DP) directly to the agent's communications~\cite{patel2022model}, adding calibrated noise to provide guarantees that an explanation does not leak significant information about an individual's data.





\subsection{The Interface and Interaction Challenge: Designing for Usable Communication}

Even when an agent is aligned to communicate ethically and transparently, its effectiveness ultimately depends on the design of the user interface. Privacy-related interactions should be efficient and minimally intrusive. If communication is delivered through frequent pop-up dialogs or interruptions, users are likely to experience notification fatigue, prompting them to ignore messages or default to permissive settings. This presents a core challenge in designing adaptive and context-aware interfaces that are resilient against manipulation. To address this, an agent should learn to trigger explicit communication only when necessary, such as during novel or high-stakes actions where its belief about user preferences is highly uncertain, while otherwise acting on its learned policy. Furthermore, the interface should use context-aware modalities, handling low-risk requests with subtle notifications while reserving explicit conversational interactions for high-risk decisions. Rather than presenting a binary choice, the interface could empower users by offering a curated set of options along the privacy-utility Pareto frontier, allowing them to make a nuanced trade-off by selecting from distinct, optimal alternatives~\cite{zhang2024adanonymizer}. Such a design carefully considers visual hierarchy, language, and defaults to empower users rather than coerce them through common dark patterns~\cite{gerl2020let}.

\subsection{The Trust and Adoption Challenge: Building a Trustworthy Communicator}

The success of our alignment framework hinges on user trust in the AI agent, especially when the agent requests access to personal data. Trust is a one-time grant; it is a fragile, cumulative state built through consistent, transparent interactions. User trust in AI is consistently built on its perceived competence, reliability, and ethical behavior in acting in the user's best interest~\cite{jacovi2021formalizing}. Each interaction effectively tests the agent’s trustworthiness, and a single breach, such as hiding information or violating a negotiated boundary, may irreparably damage that trust. To address this core challenge, future work should focus on two directions. First, we must pursue demonstrable faithfulness by creating auditable systems where users can inspect the agent's learned preference model ($R_{user}$) to verify that its actions align with their values. Second, the challenges of alignment, communication, interface design, and trust are deeply intertwined and cannot be addressed in isolation. Progress requires a holistic, human-centered approach that co-designs the alignment process, communication protocols, and interaction mechanisms as an integrated system. This ensures that improvements in one area, such as clearer explanations, do not inadvertently introduce problems in another, such as increased privacy risks.

\subsection{Privacy of Alignment}

A fundamental challenge arises from the learning requirements of our framework: aligning an agent with a user's privacy preferences necessitates collecting and modeling those very preferences, which are themselves highly sensitive. This creates a central paradox, where the process designed to enhance privacy also introduces a new potential vector for privacy compromise.

To mitigate this risk, we advocate for a multi-layered technical strategy that prioritizes data minimization and secure computation. Primarily, we advocate for the model to adopt localized processing~\cite{zhou2025rescriber}, where it learns and refines a user's preference model directly on their personal device, thereby keeping sensitive information within the user's control. When model improvements require learning from multiple users, we recommend employing privacy-preserving machine learning techniques such as federated learning combined with differential privacy. This approach allows a global model to benefit from user interactions without centralizing the sensitive data itself. Furthermore, for any data that must be transmitted off-device, we argue for the adoption of robust, formally defined anonymization and encryption protocols to ensure the security and integrity of the information while in transit.

\section{Discussions}
\label{sec:discussion}
In this section, we situate our proposed alignment framework within a broad context, discussing the socio-technical factors governing its adoption, the nuanced and dynamic nature of privacy it seeks to address, and its potential for generalization beyond simple trade-offs. We conclude by examining its practical considerations and regulatory compliance.


\subsection{Trust, Awareness and Adoption}

Effective user adoption of AI agents for privacy control is critically dependent on establishing and maintaining user trust, which in turn facilitates the user's willingness to articulate their privacy preferences and intentions. The initial perception of an AI agent's trustworthiness can be influenced by its mode of deployment. When an AI agent is pre-installed as a native feature on a device, such as a smartphone or personal computer (e.g., Apple's Siri or Microsoft's Cortana), a foundational level of trust is often implicitly transferred from the device manufacturer or operating system provider. This corporate endorsement can foster confidence in the agent's on-device processing capabilities and its adherence to established privacy norms.

Conversely, when AI agents are integrated as features within broader AI models or agent frameworks released by enterprises, it becomes imperative for the developing entity to proactively cultivate trust through transparent communication. This involves explicitly declaring the agent's on-device processing nature, if applicable, and providing detailed, comprehensible explanations of the local data processing logic. Such transparency helps to mitigate potential user mistrust regarding data handling and reinforces the agent's perceived role of user information, thereby encouraging open and honest feedback for effective privacy alignment.

AI agents should also adhere to social norms by proactively informing users of privacy risks they may not be aware of. Users often lack a comprehensive understanding of all potential privacy risks, which impedes their ability to make effective choices. This is substantiated by prior research demonstrating that users who are unaware of memory-related privacy risks are less likely to exercise control over them~\cite{zhang2024ghost}. Furthermore, pioneering studies have explored methods for reminding users about the risks of geolocation inference from their social media data~\cite{ma2025privacy}. We contend that such reminders would significantly enhance the effectiveness of alignment and mitigate user privacy fatigue~\cite{choi2018role}. Nevertheless, existing literature also indicates that communicating privacy risks and settings to users, particularly through specialized methods, presents considerable challenges, as users may not always achieve accurate comprehension~\cite{franzen2022private}. Consequently, this area requires further investigation in future work.

The real-world adoption of the alignment depends on stakeholders' adoption in diverse technological and social contexts. Transitioning from static controls requires more than technical validation; it demands a coordinated effort to address the concerns of users, developers, and regulators while proving its practical value.

For users, adoption hinges on trust built through competent, reliable, and ethical agent behavior. Our method empowers users by shifting them from passive operators to strategic directors, verifiers, and teachers, which can alleviate the cognitive burden of traditional privacy management. To prevent ``negotiation fatigue'', implementation must be seamless and context-aware, using subtle notifications for low-risk requests and explicit dialogues for high-stakes decisions.

For developers and businesses, our method offers a solution to the untenable privacy-utility trade-off, providing a competitive advantage by building trustworthy services instead of relying on fragile ``notice-and-consent'' models. However, industry adoption requires overcoming the challenge of designing reward functions that incentivize ``honest negotiation'' and prevent agents from evolving into manipulative ``dark pattern'' generators.

\subsection{Contextual and Implicit Nature of Privacy}

According to the theory of CI~\cite{nissenbaum2004privacy}, privacy risks and harms are contingent upon the type of data collected, the context of collection, and the entities involved in data transmission and sharing. Concurrently, as per the privacy calculus theory~\cite{laufer1977privacy}, user privacy preferences are influenced by the nature of the task and its perceived benefits. Moreover, user preferences are shaped by their awareness of different categories of privacy risks, such as the distinction between direct data leakage and inference-based risks~\cite{ma2025privacy,zhang2024ghost}. Therefore, the process of privacy alignment must comprehensively account for these factors, potentially through parametric modeling.

Our approach differs fundamentally from static inferences, such as traditional permission controls~\cite{micinski2017user}, in two primary aspects. First, static permission controls cannot accommodate the diversity of tasks and contexts, often preventing users from achieving an optimal trade-off between privacy and usability. Second, these static systems typically require users to engage in manual configuration, a process that has been shown to be exceedingly cumbersome~\cite{fang2014permission}.

Furthermore, our method is distinct from explicit preference modeling techniques~\cite{yang2024feasibility}. The key distinctions are threefold. Initially, users often find it difficult to articulate their privacy preferences explicitly. Additionally, the process of formally modeling these preferences can be intrusive~\cite{asthana2024know}. Finally, because these preferences are intrinsically linked to specific user tasks, repeatedly performing explicit modeling and privacy calculus for each new task would be prohibitively complex.

In our framework, we delegate the privacy calculus process to the agent. This choice reflects a practical challenge: in conversational recommendation settings, users often cannot directly perceive performance changes without first seeing the response. However, for agents, especially for LLMs, estimating the privacy, the utility, and optimizing trade-offs may be feasible given the contextual awareness of LLMs~\cite{shaoprivacylens} and the pioneering work of modeling trade-offs by LLMs~\cite{zhao2024privacy}. However, this approach raises an important question: does the agent's internal trade-off process reflect the one users would make themselves? And even if it does, how can the agent transparently communicate that process in a way that remains aligned with the user’s expectations and values? Previous work has typically qualitatively modeled humans' privacy-utility trade-off mental model through behavior analysis or interviews. Our method could adopt a similar approach, allowing agents and users to intermittently engage in reflective discussions about the trade-off process and outcomes.

\subsection{Beyond Simple Privacy-Utility Trade-offs}

In the broad landscape of privacy concerns, privacy protection often involves a trade-off with public safety. For instance, government regulation aimed at ensuring public safety may necessitate access to private information such as individuals' identification numbers and behavioral records. Such access if often deemed crucial for fostering good societal order. 

On the other hand, privacy frequently conflicts with the broader interests of the community. For example, OpenAI's use of user data to train its models can be viewed as an action that enhances collective benefit by improving the overall quality and experience of services for all users. These varying objectives can influence users' privacy choices and decisions. This observation aligns with the CI theory, which posits that individuals' privacy preferences are shaped by the context in which information is shared.

Previous research also indicates that when data is transmitted or shared with different recipients for diverse purposes, users' perceptions of the trade-off between data privacy and utility, and consequently their privacy preferences, will vary. Currently, we have simplified the complex discussions surrounding community benefits and public safety. To better integrate users' considerations regarding these aspects into our framework, future work should provide detailed explanations during privacy communications with users. This approach will allow a comprehensive understanding of users' mental models of privacy.

Our framework is grounded in a simple scenario of privacy-utility trade-offs in conversational recommendation. However, in real-world systems, particularly on smartphones or websites, privacy management is often more complex. It may involve permission systems and implicit data collection that is not directly tied to immediate utility. Feedback mechanisms can also vary: for instance, a device might reference permission settings from other apps to inform automatic configurations. Exploring how such contextual feedback influences user preferences could help extend the generalizability of our approach.


\subsection{Privacy Compliance}

While this paper primarily focuses on aligning AI agents with user privacy-task trade-off preferences, ensuring compliance with legal frameworks such as the General Data Protection Regulation (GDPR) is essential. This includes adhering to principles like data minimization~\cite{sharma2024m} and providing clear information regarding the scope of data transmission and sharing~\cite{lee2024priviaware}. Previous work has extensively explored these aspects through various user interfaces and algorithmic solutions~\cite{sharma2024m,lee2024priviaware}. For the alignment of AI agents, these compliance requirements can be framed as additional constraints. Such an approach could facilitate the direct adoption of our proposed alignment techniques by enterprises.

\section{A Call for a Human-Centered Research Agenda}
\label{sec:actions}
To address the foundational challenges and move toward a future of trustworthy AI agents, we call for an interdisciplinary and human-centered research agenda. This agenda focuses on building the theoretical, technical, and empirical foundations needed to make dynamic privacy alignment a reality, with a particular focus on actionable research for the usable privacy and security community.

\subsection{Action 1: Develop Efficient and Usable Preference Elicitation Techniques}

A core premise of our framework is the agent's ability to learn a user's latent privacy preferences through interaction. However, the CIRL process can be data-intensive, and with the traditional RLHF paradigm, there is a significant risk of imposing an excessive cognitive burden on the user, leading to ``negotiation fatigue''. A key research priority, therefore, is to develop preference elicitation techniques that are both information-efficient and cognitively lightweight. This requires moving beyond simple, repetitive queries. Research could investigate mixed-initiative interaction models where the agent uses active learning principles to determine the most informative questions to ask~\cite{horvitz1999principles}, minimizing the number of explicit negotiations. Furthermore, work is needed to develop robust methods for interpreting the rich, implicit signals present in user behavior, such as rephrased queries, ignored suggestions or interaction timings, as these can serves as powerful, low-effort sources of feedback to continuously refine the agent's model of the user's privacy calculus.

\subsection{Action 2: Design and Evaluate Privacy Communication Interfaces}

Even with an efficient learning strategy, the success of this framework depends on the interface through which privacy negotiation occurs. An intrusive or poorly designed interface will lead users to ignore requests or abandon the tool altogether. Thus, we call for design-led research focused on building and empirically evaluating adaptive communication interfaces. This work should explore how interface modality and intrusiveness can be dynamically tailored to match the risk level of each interaction. For instance, a low-risk negotiation, such as using session data to clarify an ambiguous request, might be handled by a subtle, dismissible notification. A high-risk request, like sharing sensitive information with a new third party, should require a more explicit conversational interaction. A further design goal is to build manipulation-resilient interfaces that frame choices in a neutral and balanced way, avoiding coercive patterns or deceptive framing tactics.

\subsection{Action 3: Establish Benchmarks for Measuring Privacy Alignment and Trust}

The effectiveness of privacy alignment cannot be evaluated using conventional AI metrics such as accuracy or task success alone. A perfectly functional agent could be untrustworthy even if it achieves its goals through manipulation or by violating user expectations. Therefore, we call for the creation of new, human-centered benchmarks designed specifically to evaluate the quality and trustworthiness of privacy-aware agents from multiple dimensions~\cite{shaoprivacylens}. These metrics should reflect the nuanced success of privacy alignment, including validated scales to measure perceived transparency, trust, and privacy protection, as well as the longitudinal performance of the alignment process.

\begin{acks}
This work was supported by the Natural Science Foundation of China under Grant No. 62472243 and 62132010.
\end{acks}

\bibliographystyle{ACM-Reference-Format}
\bibliography{sample-base}

\appendix

\end{document}